\documentclass[conference]{IEEEtran}
\IEEEoverridecommandlockouts
\usepackage{float}
\usepackage{comment}
\usepackage{multirow}
\usepackage{cite}
\usepackage{url}
\usepackage{amsmath,amssymb,amsfonts}
\usepackage{algorithmic}
\usepackage{graphicx}
\usepackage{textcomp}
\usepackage{xcolor}
\def\BibTeX{{\rm B\kern-.05em{\sc i\kern-.025em b}\kern-.08em
    T\kern-.1667em\lower.7ex\hbox{E}\kern-.125emX}}

\begin{document}

\title{Scalable Quantum-Classical DFT Embedding for NISQ Molecular Simulation}

\author{

\IEEEauthorblockN{
Namrata Manglani\IEEEauthorrefmark{1}\IEEEauthorrefmark{2},
Samrit Kumar Maity\IEEEauthorrefmark{1},
Ranjit Thapa\IEEEauthorrefmark{3},
Sanjay Wandhekar\IEEEauthorrefmark{1}
}

\IEEEauthorblockA{\IEEEauthorrefmark{1}
C-DAC, Pune, India
}

\IEEEauthorblockA{\IEEEauthorrefmark{2}
Shah and Anchor Kutchhi Engineering College, Mumbai, India
}

\IEEEauthorblockA{\IEEEauthorrefmark{3}
SRM University-AP, Amaravati, India
}

\IEEEauthorblockA{
Email: namrata.manglani@sakec.ac.in, samritm@cdac.in, ranjit.t@srmap.edu.in, sanjayw@cdac.in
}

}

\maketitle

\begin{abstract}
Scalable quantum–classical embedding is essential for chemically meaningful simulations on near-term NISQ hardware. Using QDFT, we show systematic correlation energy recovered relative to the baseline DFT, measured against CCSD in a fixed six-orbital active space across molecules from water to naphthalene. By varying embedded electrons from 2 to 8, aromatic systems saturate near 63 to 64\%, while linear molecules such as carbon dioxide reach 68\%. All systems converge within two iterations of embedding under relaxed self-consistency thresholds, highlighting robustness. An active space (4e, 6o) recovers the ~60\% correlation using 10 qubits, providing practical NISQ guidelines.
\end{abstract}


\section{Introduction}
\label{sec:intro}

Accurate electronic structure simulations require correlation beyond mean-field DFT, yet computational demand of exact methods like full configuration interaction scales factorially with system size. While Coupled Cluster Methods CCSD and CCSD(T) provide practical alternatives, their cost limits applications to $\sim$50 electrons~\cite{Bartlett2007}. Near-term NISQ devices offer acceleration via VQE~\cite{Peruzzo2014}, but qubit constraints restrict active space sizes~\cite{McArdle2020}.

Quantum embedding methods address this challenge by partitioning a molecular system into a strongly correlated active region treated with high-level theories and an environment described using more efficient approximations. Fragment-based approaches such as Density Matrix Embedding Theory (DMET)~\cite{Knizia2012} typically require large impurity regions, whereas projection-based QDFT embedding~\cite{Rossmannek2021} enables the use of minimal active spaces (here, six orbitals) within a DFT bath. This framework captures dynamic correlation at the classical level while restricting quantum computation to compact regions well suited for NISQ hardware.

We advance this framework with two key contributions: (1) systematic active-space scaling across chemically diverse molecules (water to naphthalene), revealing aromatic saturation at 63–64\% CCSD correlation recovery (6e,6o) versus linear carbon dioxide scaling to 68\% (8e,6o); and (2) robust 2-iteration SCF convergence across 3–18 atom systems via adaptive density damping. These establish production-ready NISQ guidelines using $\sim$10 qubits, with naphthalene as the largest QDFT embedding demonstration to date.

\section{Computational Methods}
\label{sec:methods}

\subsection{Quantum DFT Embedding Framework}

All calculations were performed using a range-separated density functional theory (DFT) embedding formalism following Rossmannek et al.~\cite{Rossmannek2021}. The calculations were carried out using Qiskit Nature version 0.7.2~\cite{QiskitNature} interfaced with PySCF version 2.6.2~\cite{PySCF2018}. The embedding workflow was implemented using a custom \texttt{DFTEmbeddingSolver} code provided in Ref.~\cite{Manglani2026}, which enables reproducibility with currently supported software versions.

The embedding bath was treated using a range-separated Local Density Approximation (LDA)~\cite{KohnSham1965} with long-range Hartree--Fock exchange. The baseline DFT calculations employed the LDA with the Vosko--Wilk--Nusair (VWN) correlation functional~\cite{Vosko1980}. The 6-31G* basis set was used consistently for the active region, embedding bath, and reference calculations.

Relative to the original Qiskit AquaChemistry prototype~\cite{RossmannekGitHub}, the present implementation incorporates three numerical stabilization advances: (i) Hartree–Fock initialization of the VQE solver with variational parameters from a Gaussian perturbation ($\sigma = 10^{-3}$) to break parameter symmetries; (ii) adaptive density damping in the embedding cycle, $\alpha_i = \max(0.05, 0.2/\sqrt{i})$; and (iii) a relaxed embedding convergence threshold of $10^{-7}$ Ha~\cite{Pes2023,Foerster2021}.

\begin{table}[htbp]
\centering
\caption{Water(2e,6o) parameter sensitivity analysis. All variations maintain 2-iteration convergence.}
\label{tab:sensitivity}

\begin{tabular}{@{}lcccc@{}}
\hline
\textbf{Parameter} & \textbf{Variation} & $\Delta E_\text{total}$ & $\Delta E_\text{active}$ & \textbf{Iterations} \\
\hline

\multirow{3}{*}{Damping $\alpha_\text{init}$}
 & 0.2 (ref) & 0.00 $\mu$Ha & 0.00 $\mu$Ha & 2 \\
 & 0.24 (+20\%) & 0.00 $\mu$Ha & +0.95 $\mu$Ha & 2 \\
 & 0.16 (-20\%) & 0.00 $\mu$Ha & -0.94 $\mu$Ha & 2 \\
\hline

\multirow{4}{*}{VQE $\sigma$}
 & 0.001 (ref) & 0.00 $\mu$Ha & 0.00 $\mu$Ha & 2 \\
 & 0.01 (10x) & -0.20 $\mu$Ha & -0.06 $\mu$Ha & 2 \\
 & 0.1 (100x) & +0.26 $\mu$Ha & -0.01 $\mu$Ha & 2 \\
 & zeros & -0.07 $\mu$Ha & 0.00 $\mu$Ha & 2 \\
\hline

\end{tabular}
\end{table}

Table~\ref{tab:sensitivity} confirms parameter robustness for water: $\pm$20\% damping variations show 0.00 µHa total energy deviation, while VQE initialization tolerates 100x $\sigma$ changes ($\leq 0.26$ µHa). These empirical parameters achieve 2-iteration convergence across all studied systems (3–18 atoms).

\subsection{Range-Separation Parameter Optimization}

The range-separation parameter $\mu$ was optimized individually for each molecular
system by minimizing the fully converged quantum DFT embedding energy. For all
systems, $\mu$ was scanned from 0.5 to 10 in increments of 0.25, covering the
range from near the pure DFT limit ($\mu \to 0$) toward the Hartree--Fock limit
($\mu \to \infty$).

At each $\mu$, the embedding equations were iterated to full self-consistency using
identical active-space definitions, quantum solvers, and convergence thresholds.
Only fully converged embedding solutions were used to evaluate the total
embedding energy $E_{\mathrm{QDFT}}$. The optimal $\mu$ was then chosen as the
value that yielded the lowest $E_{\mathrm{QDFT}}$, and this $\mu_{\mathrm{opt}}$
was held fixed for all subsequent calculations on the corresponding molecular system.

\subsection{Active Space and Quantum Solver Details}

All embedded quantum calculations employed a fixed active space of six spatial orbitals (2e–8e,6o scaling). This compact choice balances chemical completeness with NISQ hardware constraints: it captures dominant valence correlation while delegating dynamic correlation to the DFT bath. Larger $\pi$-spaces like CAS(10,10) exceed current qubit limits for aromatic systems such as naphthalene.

The Second-quantized Hamiltonians used parity mapping~\cite{Seeley2012parity} combined with symmetry-based qubit
tapering~\cite{Bravyi2017tapering} to reduce the required qubit count. The UCCSD ansatz was initialized from a Hartree–Fock (HF) reference state and implemented using a single Trotter step. Variational parameters were initialized from a seeded Gaussian distribution with standard deviation $\sigma = 0.001$, ensuring reproducibility and keeping the ansatz close to the HF reference, which improves local curvature estimation and stabilizes convergence. VQE optimization was performed using L-BFGS-B~\cite{Byrd1995} with a maximum of 500 iterations and an energy convergence tolerance of $10^{-6}$. A callback was used to record energy evaluations during optimization, and all expectation values were computed using an exact (noiseless) quantum estimator, isolating embedding and algorithmic performance from hardware noise effects.

\subsection{Molecular Systems and Reference Data}

\begin{figure}[htbp]
\centering
\includegraphics[width=0.5\textwidth,draft=false]{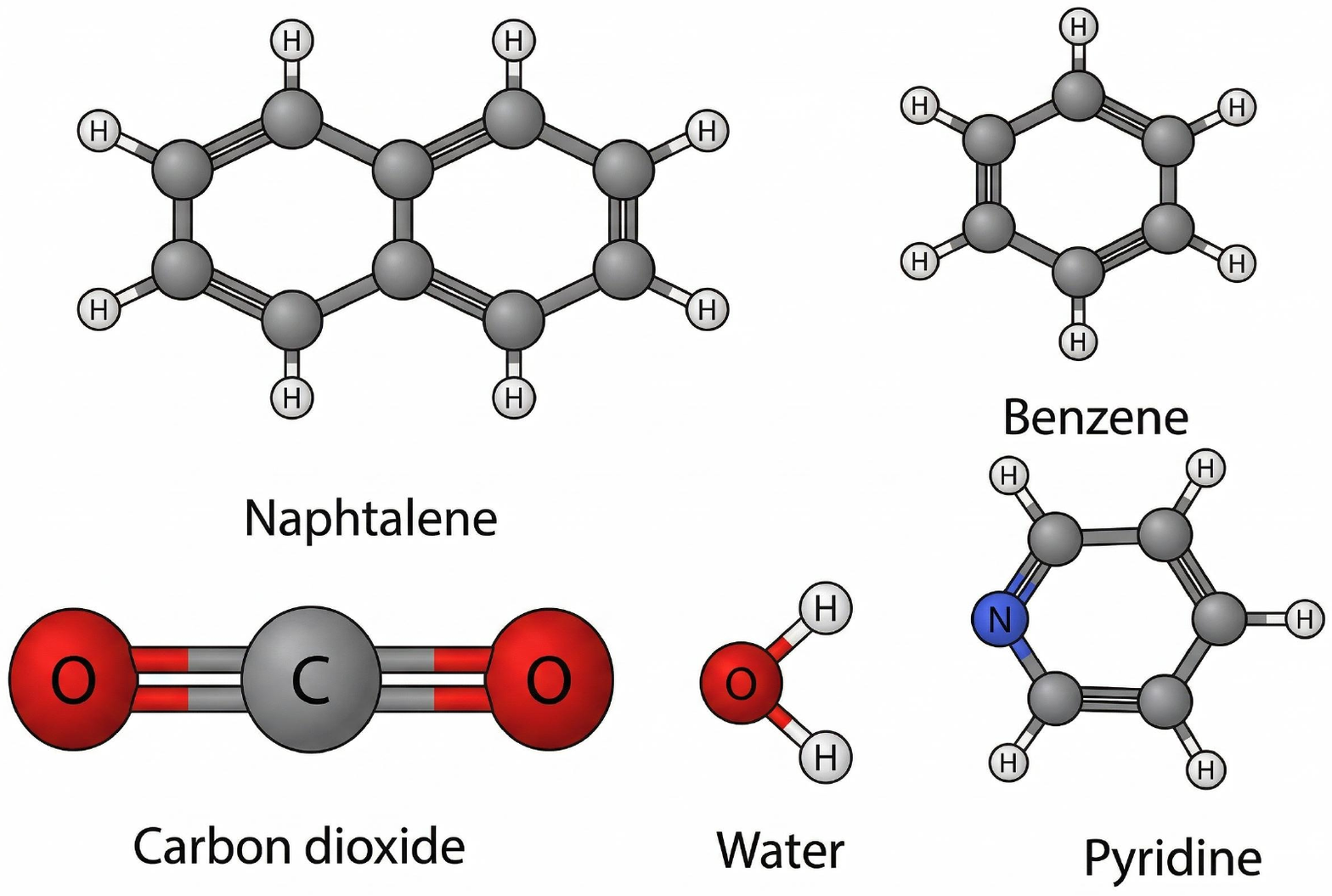}

\caption{Molecular structure representation derived from MolView input and visually refined using Gemini; the structure was verified by the authors}
\label{fig:mol}
\end{figure}

The molecular test set comprises Water (H$_2$O), Carbon dioxide (CO$_2$), Benzene (C$_6$H$_6$), Pyridine (C$_5$H$_5$N), and Naphthalene
(C$_{10}$H$_8$) as shown in Figure~\ref{fig:mol}, the latter representing the largest system studied in a quantum DFT
embedding context to date. All geometries correspond to equilibrium structures
obtained from the NIST Computational Chemistry Comparison and Benchmark Database
(CCCBDB) at the B3LYP/6-31G* level~\cite{cccbdb}.

Reference correlation energies were obtained from coupled-cluster singles and doubles
(CCSD) calculations performed in \texttt{PySCF}, using identical geometries and basis
sets as employed in the embedding calculations.

\section{Results and Discussion}
\label{sec:results}

\subsection{Baseline Energies}
Table~\ref{tab:baselines} reports total energies obtained at the Hartree--Fock (HF), 
density functional theory (DFT), quantum DFT embedding with a VQE active-space solver 
(QDFT), and CCSD levels for all systems considered, using a consistent 6-31G* basis 
and equilibrium geometries.

Across all molecules, QDFT systematically improves upon both HF and DFT energies and 
approaches the CCSD reference. This trend is observed consistently for small polar 
systems (H$_2$O), linear molecules (CO$_2$), and conjugated aromatic systems, 
demonstrating the robustness of the embedding protocol across chemically diverse 
regimes.

\begin{table}[htbp]
\centering
\caption{Baseline total energies (Ha) for the studied molecules using the 6-31G* basis. QDFT values correspond to the best-performing active space $(e,o)$, with CCSD energies used as reference for correlation recovery.}
\label{tab:baselines}
\begin{tabular}{@{}lcccc@{}}
\hline
Molecule (atoms) & HF & DFT & QDFT (e,o) & CCSD \\
\hline
Water (3)        & -76.008 & -75.841 & -76.068 (6,6) & -76.205 \\
Carbon dioxide (3)       & -187.630 & -187.174 & -187.807 (8,6) & -188.102 \\
Benzene (12)     & -230.701 & -230.074 & -230.991 (6,6) & -231.502 \\
Pyridine (11)    & -246.693 & -246.043 & -246.980 (6,6) & -247.519 \\
Naphthalene (18) & -383.352 & -382.320 & -383.818 (6,6) & -384.674 \\
\hline
\end{tabular}
\end{table}

\subsection{Correlation Recovery and Active-Space Scaling}
The recovery fraction
\begin{equation}
R = 100 \times 
\frac{E_{\mathrm{QDFT}} - E_{\mathrm{DFT}}}
     {E_{\mathrm{CCSD}} - E_{\mathrm{DFT}}}
\end{equation}
quantifies the fraction of correlation energy recovered by QDFT.
Here, $E_{\mathrm{DFT}}$ is the total energy from an Restricted Kohn Sham(RKS) PySCF
calculation using the 6-31G* basis set and the LDA (VWN) functional,
$E_{\mathrm{QDFT}}$ is the corresponding energy obtained with quantum
embedding applied to the same DFT description, and $E_{\mathrm{CCSD}}$
is the coupled-cluster singles and doubles reference energy. A value
of $R=0\%$ corresponds to plain DFT, while $R=100\%$ indicates full
recovery of the CCSD correlation energy.

Table~\ref{tab:recovery} summarizes the maximum correlation recovery obtained for fixed 
six-orbital active spaces while systematically increasing the number of correlated 
electrons. All systems exceed 60\% recovery at modest active-space sizes (Results for water and pyridine can be compared with \cite{Rossmannek2021}), with carbon dioxide 
reaching a maximum of 68.0\% at (8e,6o). Aromatic systems exhibit a clear plateau, saturating near 63--64\% recovery at (6e,6o).

\begin{table}[htbp]
\centering
\caption{CCSD correlation recovery (\%) for fixed six-orbital active spaces as a function of active-electron count. $\mu_{\mathrm{opt}}$ denotes the optimal range-separation parameter used in the embedding calculation.}
\label{tab:recovery}
\begin{tabular}{@{}lccccc@{}}
\hline
Molecule & $\mu_{\mathrm{opt}}$ & (2e,6o) & (4e,6o) & (6e,6o) & (8e,6o) \\
\hline
Water        & 7.25 & 61.8 & 60.1 & \textbf{62.1} & 61.4 \\
Carbon dioxide & 6.75 & 63.7 & 66.4 & 66.9 & \textbf{68.2} \\
Benzene      & 5.00 & 62.6 & \textbf{64.2} & 64.2 & 64.2 \\
Pyridine     & 5.25 & 62.0 & \textbf{63.5} & 63.5 & 63.5 \\
Naphthalene  & 5.00 & 62.2 & 62.8 & \textbf{63.7} & 63.0 \\
\hline
\end{tabular}
\end{table}

\subsection{Active-Space Convergence Trends}
\begin{figure}[htbp]
\centering
\includegraphics[width=\columnwidth]{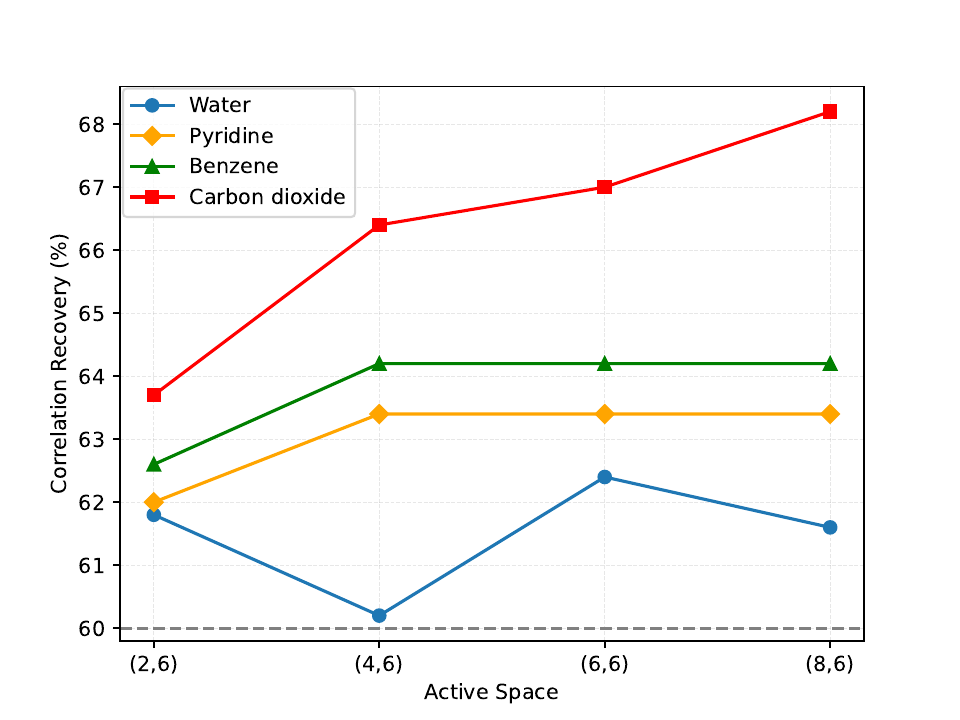}   
\caption{Correlation recovery versus active-space size. The dashed line indicates 60\% 
recovery. Linear CO$_2$ continues to scale, while aromatic systems saturate at modest 
active spaces.}
\label{fig:convergence}
\end{figure}

Figure~\ref{fig:convergence} illustrates correlation recovery as a function of active-space 
electron count for all systems. A 60\% recovery threshold (dashed line) is exceeded in 
all cases once four or more electrons are correlated, demonstrating efficient 
convergence within small active spaces.

Carbon dioxide exhibits the most favorable scaling behavior, continuing to gain correlation 
energy up to (8e,6o), whereas aromatic systems plateau at approximately 63--64\% recovery. 
This behavior is consistent with the rapid saturation of $\pi$-correlation in cyclic 
conjugated systems reported in multireference studies~\cite{Roos2011,Celani1995} and in 
quantum embedding contexts~\cite{Rossmannek2021,Battaglia2024}.

\subsection{Discussion}

The central outcome of this work is that QDFT provides systematic and size-consistent improvements over classical DFT while maintaining a fixed quantum computational cost as molecular size increases. Fixed six-orbital active spaces recover more than 60\% of the CCSD correlation energy across chemically diverse systems (Table~\ref{tab:recovery}), requiring only $\sim$10 qubits after symmetry tapering. This establishes a practical operating regime for NISQ hardware and highlights a key advantage of embedding approaches: quantum resource requirements remain effectively decoupled from overall system size~\cite{Battaglia2024}.

In contrast to fragment-based embedding methods such as density matrix embedding theory (DMET)~\cite{Knizia2012}, which often require larger impurity regions to capture extended correlation effects, the QDFT framework leverages the DFT bath to recover dynamic correlation outside the quantum region. This enables chemically meaningful correlation trends to be reproduced using compact active spaces compatible with near-term quantum hardware.

Correlation recovery trends reflect underlying molecular structure. Aromatic systems exhibit early saturation near 63--64\%, consistent with rapid convergence of dominant $\pi$-electron correlations once key near-degeneracies are included. In contrast, the linear molecule carbon dioxide continues to benefit from increased active-space correlation, reaching up to 68\% recovery at the (8e,6o) level.

The embedding procedure is numerically robust across all systems, with convergence typically achieved within two iterations using a relaxed SCF threshold of $10^{-7},\mathrm{Ha}$. Sensitivity analysis (Table~\ref{tab:sensitivity}) shows minimal variation (0.00--0.26 $\mu$Ha), supported by small random VQE initialization and adaptive density damping, indicating stable and reliable embedding behavior.

\noindent \textbf{NISQ Considerations:} The present results employ noiseless exact estimators to isolate embedding performance from hardware effects. Transition to realistic NISQ devices will require noise-resilient optimization strategies such as Simultaneous Perturbation Stochastic Approximation (SPSA)~\cite{Spall1992}, along with error mitigation techniques including zero-noise extrapolation (ZNE) and readout correction. Preliminary noisy simulations using AER indicate that SPSA remains viable under noise, whereas gradient-based optimizers such as L-BFGS-B fail to converge reliably. Experimental validation on IBM Quantum hardware is planned as a next step.
\section{Conclusions}
\label{sec:conclusions}

Range-separated DFT embedding with a quantum active-space solver recovers 60--68\% of CCSD correlation energy using compact active spaces (up to 8 electrons in 6 orbitals), demonstrated across water, carbon dioxide, benzene, pyridine, and naphthalene (6-31G*, LDA) [Tables~\ref{tab:baselines}--\ref{tab:recovery}].

Two key advances enable this performance: (1) fixed active spaces that decouple quantum computational cost from molecular size (3--18 atoms), and (2) robust convergence achieved through small random VQE initialization and adaptive density damping, requiring only two embedding iterations (Table~\ref{tab:sensitivity}).

Correlation recovery trends demonstrate aromatic systems saturate efficiently near 63--64\%, consistent with rapid convergence of $\pi$-electron correlations, while linear CO$_2$ continues to scale, reaching 68\% at the (8e,6o) level (Fig.~\ref{fig:convergence}).

\noindent \textbf{Limitations:} The present study employs noiseless exact estimators to isolate embedding performance. While this clarifies algorithmic behavior, practical deployment on NISQ hardware will require addressing noise, optimizer stability, and measurement overhead. The relaxed $10^{-7}$ Ha convergence threshold further motivates validation under realistic noise conditions.

\noindent \textbf{Future Work:} Immediate next steps include execution on quantum hardware (e.g., IBM Quantum) using noise-resilient optimizers such as SPSA, combined with error mitigation techniques including zero-noise extrapolation (ZNE) and readout correction. Additional directions include exploring higher-level DFT functionals in the bath and extending the framework to strongly multireference and excited-state systems.

Overall, this work establishes quantum DFT embedding as a scalable, chemically interpretable framework that bridges near-term quantum algorithms and realistic molecular applications.

\section*{Acknowledgment}

The author(s) gratefully acknowledge the support of the AICTE Industry Fellowship Scheme for funding and facilitating this research. The guidance and resources provided under this program have been invaluable in carrying out this work. The author(s) also thank the Centre for Development of Advanced Computing (C-DAC) and the National Supercomputing Mission (NSM) for providing computational resources and technical support, which significantly contributed to the completion of this research.

\IEEEtriggeratref{4}

\end{document}